\documentclass[runningheads]{llncs}
\usepackage{amsmath,graphicx}
\usepackage[T1]{fontenc}
\usepackage{graphicx}
\usepackage{hyperref}
\usepackage{enumitem}
\usepackage[linesnumbered,ruled,vlined]{algorithm2e}
\setlist{nosep, leftmargin=14pt}
\SetKwComment{Comment}{/* }{ */}
\usepackage{mwe} 
\RestyleAlgo{ruled}

\SetCommentSty{mycommfont}
\SetAlFnt{\small\rmfamily}
\usepackage{booktabs}
\usepackage[table,xcdraw]{xcolor}
\usepackage{graphicx}
\usepackage{subcaption}
\begin{document}
	\title{Infusing physically inspired known operators in deep models of ultrasound elastography }
	\titlerunning{Known operators for ultrasound elastography}
	%
	\author{Ali K. Z. Tehrani$^1$ and Hassan Rivaz$^2$}
	%
	\authorrunning{Ali K. Z. Tehrani}
	%
	%
	\institute{$^{1,2}$Department of Electrical and Computer Engineering, Concordia University, Canada \\
		$^1$\email{a\_kafaei@encs.concordia.ca} $^2$\email{hrivaz@ece.concordia.ca}\\}
	\maketitle              
	\begin{abstract}
		The displacement estimation step of Ultrasound Elastography (USE) can be done by optical flow Convolutional Neural Networks (CNN). Even though displacement estimation in USE and computer vision share some challenges, USE displacement estimation has two distinct characteristics that set it apart from the computer vision counterpart: high-frequency nature of RF data, and the physical rules that govern the motion pattern. The high-frequency nature of RF data has been well addressed in recent works by modifying the architecture of the available optical flow CNNs. However, insufficient attention has been placed on the integration of physical laws of deformation into the displacement estimation. In USE, lateral displacement estimation, which is highly required for elasticity and Poisson's ratio imaging, is a more challenging task compared to the axial one since the motion in the lateral direction is limited, and the sampling frequency is much lower than the axial one. Recently, Physically Inspired ConstrainT for Unsupervised Regularized Elastography (PICTURE) has been introduced which incorporates the physical laws of deformation by introducing a regularized loss function. PICTURE tries to limit the range of the lateral displacement by the feasible range of Poisson's ratio and the estimated high-quality axial displacement. Despite the improvement, the regularization was only applied during the training phase. Furthermore, only a feasible range for Poisson's ratio was enforced. We exploit the concept of known operators to incorporate iterative refinement optimization methods into the network architecture so that the network is forced to remain within the physically plausible displacement manifold. The refinement optimization methods are embedded into the different pyramid levels of the network architecture to improve the estimate. Our results on experimental phantom and \textit{in vivo} data show that the proposed method substantially improves the estimated displacements.
		
	\end{abstract}
	\section{Introduction}
	\label{sec:intro}
	Ultrasound Elastography (USE) provides information related to the stiffness of the tissue. Ultrasound (US) data before and after the tissue deformation (which can be caused by an external or internal force) are collected and compared to calculate the displacement map, indicating each individual sample's relative motion. The strain is computed by taking the derivative of the displacement fields. In free-hand palpation, the force is external and applied by the operator by the probe \cite{ophir1999elastography}.  
	
	Convolutional Neural Networks (CNN) have been successfully employed for USE displacement estimation \cite{tehrani2020displacement,peng2020neural,tehrani2020real}. Unsupervised and semi-supervised training methods have been proposed, which enable the networks to use real US images for training \cite{tehrani2020semi,delaunay2020unsupervised,wei2022unsupervised}. The proposed networks have achieved high-quality axial strains. In contrast to axial strain, lateral strain, which is highly required in Poisson's ratio imaging and elasticity reconstruction, has a poor quality due to the low sampling frequency, limited motion and lack of carrier signal in the lateral direction.

	Recently, physically inspired constraint in unsupervised regularized elastography (PICTURE) has been proposed \cite{kz2022physically}. This method aims to improve lateral displacement by exploiting the high-quality axial displacement estimation and the relation between the lateral and axial strains defined by the physics of motion. Despite the substantial improvement, the regularization is only applied during the training phase. In addition, only a considerably large feasible range for Poisson's ratio was enforced, thereby providing further opportunities for the network to contravene the laws of physics.

	Known operators, introduced by Maier \textit{et al.} \cite{maier2019learning}, have been widely utilized in deep neural networks. The core idea is that some known operations (for example inversion of a matrix) are embedded inside the networks to simplify the training and improving the generalization ability of the network. The known operator can be viewed as the prior knowledge related to the physics of the problem. Maier \textit{et al.} investigated known operators in different applications such as computed tomography, magnetic resonance imaging, and vessel segmentation, and showed a substantial reduction in the maximum error bounds \cite{maier2019learning}.

	In this paper, we aim to embed two lateral displacement refinement algorithms in the CNNs to improve the lateral strains. The first algorithm limits the range of Effective Poisson's Ratio (EPR) inside the feasible range during the test time. It is important to note that in contrast to \cite{kz2022physically}, the EPR range is enforced using the regularization during the training phase and the known operators framework during the test phase; therefore, it is enforced during both training and test phases. The second algorithm employs the refinement method proposed be Gou \textit{et al.} \cite{guo2015pde} which exploits incompressibility constraint to refine the lateral displacement. The network weight and a demo code are publicly available online at \href{http://code.sonography.ai}{http://code.sonography.ai}.
	\vspace{-0.15cm}
	\section{Materials and Methods}
	\label{sec:mat}
	In this section, we first provide a brief overview of PICTURE and underlie some differences to this work. We then introduce our method for incorporating known operators into our deep model and outline our unsupervised training technique. We then present the training and test datasets and finish the section by demonstrating the network architecture.

	\subsection{PICTURE}
	Let $\varepsilon_x$ denote axial ($x=1$), lateral ($x=2$), and out-of-plane ($x=3$) strains.  Assuming linear elastic, isotropic, and homogeneous material that can move freely in the lateral direction, the lateral strain can be obtained from the axial strain and the Poisson's ratio by $\varepsilon_2 = -v\times \varepsilon$. Real tissues are inhomogeneous, and boundary conditions exist; therefore, the lateral strain cannot be directly obtained by the axial strain and the Poisson's ratio alone. In such conditions, EPR, which is defined as $v_e = \frac{-\varepsilon_{22}}{\varepsilon_{11}}$ can be employed \cite{ma2014principle}. EPR is spatially variant, and it is not equal to Poisson's ratio, particularly in the vicinity of inclusion boundaries or within inhomogeneous tissue. Its value tends to converge towards the Poisson's ratio in homogeneous regions, and it has a similar range of Poisson's ratio, i.e., between 0.2 and 0.5 \cite{mott2013limits}. In PICTURE, a regularization was defined to exploit this range and the out-of-range EPRs were penalized \cite{kz2022physically}. PICTURE loss can be obtained from the following procedure: \newline{}
	1- Detect out-of-range EPRs by:
	\begin{equation}
		M(i,j) = \begin{Bmatrix}
			0 &  & v_{emin}<\widetilde{v_e}(i,j)<v_{emax} \\
			1&  &  otherwise\\
		\end{Bmatrix}
	\end{equation}	     
	where $\widetilde{v_e}$ is the EPR obtained from the estimated displacements. $v_{emin}$ and $v_{emax}$ are two hyperparameters that specify the minimum and maximum accepted EPR values, which are assumed to be 0.1 and 0.6, respectively. \newline{}
	2- Penalize the out-of-range lateral strains using:
	\begin{equation}
		\label{eq:picture}
		\begin{gathered}
			L_{vd} = \left|  (\varepsilon_{22}+<\widetilde{v_e}>\times \mathbf{S}(\varepsilon_{11}))\right|_2 \\
			\bar{V_e} = \frac{\sum_{i,j}^{}(1-M_{(i,j)}) V_e(i,j)}{\sum_{i,j}^{}(1-M_{(i,j)})}
		\end{gathered}
	\end{equation} 
	where $<\widetilde{v_e}>$ is the average of EPR values within the feasible range. The operator $\mathbf{S}$ denotes stop gradient operation, which is employed to avoid the axial strain being affected by this regularization. It should be noted in contrast to \cite{kz2022physically} in which only out-of-range samples were contributing to the loss, in this work, all samples contribute to $L_{vd}$ to reduce the estimation bias.
	\newline{}
	3- Smoothness of EPR is considered by:
	\begin{equation}
		L_{vs} = |\frac{\partial v_e}{\partial a}|_1 + \beta \times |\frac{\partial v_e}{\partial l}|_1
	\end{equation} 
	4- PICTURE loss is defined as $L_V = L_{vd} + \lambda_{vs} \times L_{vs}$, where $\lambda_{vs}$ is the weight of the smoothness loss. PICTURE loss is added to the data and smoothness losses of unsupervised training.

	\subsection{Known Operators}
	The known operators are added to the network in the inference mode only due to the high computational complexity of unsupervised training (outlined in the next section). We employ two known operators to impose physically known constraints on the lateral displacement. 
	
	The first known operator (we refer to it as Poisson's ratio clipper) limits the EPR to the feasible range of $v_{emin} - v_{emax}$. Although PICTURE tries to move all EPR values to the feasible range, in \cite{kz2022physically}, it was shown that some samples in test time were still outside of the feasible range. Poisson's ratio clipper is an iterative algorithm since the lateral strains are altered by clipping the EPR values and affecting the neighbor samples' strain values. 
	
\begin{figure}
		
		\centering
		\includegraphics[width=0.66\textwidth]{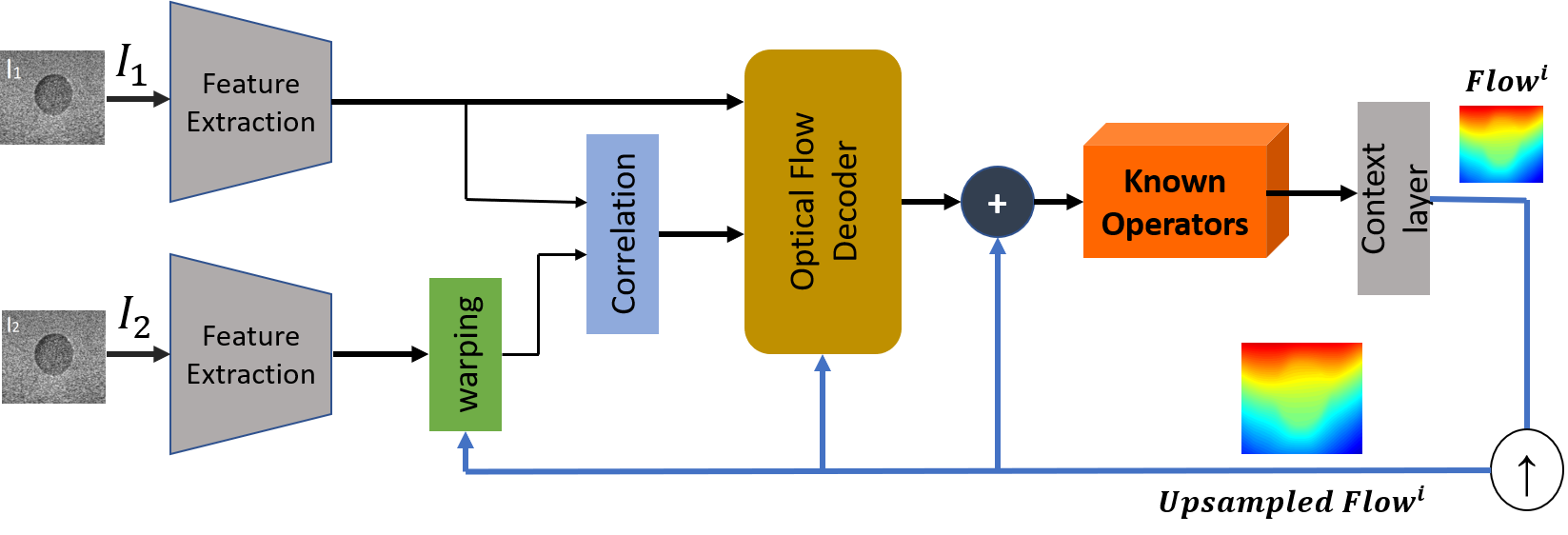}
		\caption{MPWC-Net++ architecture with known operators. The network is iterative with 5 pyramid levels. The known operators are added after optical flow estimation, and refine the estimated lateral displacement in each pyramid level (added from level 3) to provide improved lateral displacement to the next pyramid level.}
		\label{fig:net}	
	\end{figure}

	The second algorithm employs the incompressibility of the tissue which can be formulated by:
	\begin{equation}
		\label{eq:incomp}
		\varepsilon_1 + \varepsilon_2 + \varepsilon_3 = 0
	\end{equation} 
	In free-hand palpation, the force is approximately uniaxial ($\varepsilon_3 \simeq  \varepsilon_2$); therefore Eq \ref{eq:incomp} can be written as:
	\begin{equation}
		\varepsilon_1 + 2\times \varepsilon_2  = 0
	\end{equation}
	Guo \textit{et al.} enforced incompressibility in an iterative algorithm \cite{guo2015pde}. We made a few changes to increase the method's robustness by adding Gaussian filtering and using a hyper-parameter weight in each iteration. It should be noted that the algorithm can be employed for compressible tissues as well, and the incompressibility constraint is employed for the refinement of the obtained displacement. The proposed algorithms are outlined in Algorithm \ref{alg:one} and \ref{alg:two}. The network architecture with the known operators is illustrated in Fig. \ref{fig:net}. It is worth highlighting that known operators offer a compelling alternative to regularization. While the latter involves adjusting trained weights based on the training data and keeping them fixed during testing, the former relies on iterative refinement that is adaptable to the test data and does not require any learnable weights.
	
	\begin{algorithm}
		\caption{Poisson's ratio clipper}\label{alg:one}
		\SetKwData{Left}{left}\SetKwData{This}{this}\SetKwData{Up}{up}
		\SetKwFunction{Union}{Union}\SetKwFunction{FindCompress}{FindCompress}
		\SetKwInOut{Input}{input}\SetKwInOut{Output}{output}
		\Input{Lateral displacement $w_l$, axial displacement $w_a$, $v_{emin}$,$v_{emax}$, $iteration$}
		\Output{Refined lateral displacement $w_{ref}$}
		$w_{ref} \leftarrow w_l$ \\
		\For{$q\leftarrow 1$ \KwTo $iteration$}{
			$e_{22}\leftarrow\frac{\partial w_l}{\partial l}$ \tcp*[f]{gradient in lateral direction.}\\
			$e_{11}\leftarrow\frac{\partial w_a}{\partial a}$ \tcp*[f]{gradient in axial direction.}\\
			$epr \leftarrow \frac{-e_{22}}{e_{11}}$\\
			$epr(epr<v_{emin}) \leftarrow v_{emin}$\tcp*[f]{Clip epr less than $v_{emin}$}\\
			$epr(epr>v_{emax}) \leftarrow v_{emax}$\tcp*[f]{Clip epr less than $v_{emax}$}\\
			
			$w_{ref}(:,2 \:\:  to \:\:  end) \leftarrow w_{ref}(:,1 \: \: to\: \:  end-1) + epr\times e_{11}$ \\
			\tcp*[f]{use the displacement of previous line and the clipped epr to find the displacement of the next line}\\

		}
	\end{algorithm}
	
	\begin{algorithm}[t]
		\caption{Guo \textit{et al.} refinement \cite{guo2015pde} employed as known operator}\label{alg:two}
		\SetKwData{Left}{left}\SetKwData{This}{this}\SetKwData{Up}{up}
		\SetKwFunction{Union}{Union}\SetKwFunction{FindCompress}{FindCompress}
		\SetKwInOut{Input}{input}\SetKwInOut{Output}{output}
		\Input{Lateral displacement $w_l$, Axial displacement $w_a$ of size $w \times h$, $iteration$, $\lambda_1$, $\lambda_2$}
		\Output{Refined lateral displacement $w_{ref}$}
		$w_{ref} \leftarrow w_l$ \\
		\For{$q\leftarrow 1$ \KwTo $iteration$}{
			\For{$i,j$ in $w,h$}{
				$\delta = W_l(i,j-1) -2W_l(i,j) + W_l(i,j+1) +W_a(i+1,j+1) - W_a(i-1,j) - W_a(i,j-1)+W_a(i-1,j-1)+ \lambda_1(W_{ref}^{q-1}-W_{ref}^{q-2})$	\\
				$w_{ref}^q = Gauss(w_{ref}^{q-1} +\lambda_2\times \delta)$ \tcp*[f]{Gaussian filtering to reduce noise, $\lambda_2$ controls the weight of updating $w_{ref}^q$}\\		
		}}
	\end{algorithm}

	\subsection{Unsupervised Training}
	We followed a similar unsupervised training approach presented in \cite{kz2022physically} for both PICTURE and kPICTURE methods. The loss function can be written as:
	\begin{equation}
		\label{eq:total_loss}
		Loss = L_D + \lambda_S L_S + \lambda_V L_V 
	\end{equation} 
	where $L_D$ denotes photometric loss which is obtained by comparing the pre-compressed and warped compressed RF data, $L_S$ is smoothness loss in both axial and lateral directions. $\lambda_S$ and $\lambda_V$ specify the weights of the smoothness loss and PICTURE loss, respectively.
	\subsection{Dataset and Quantitative Metrics}
	We use publicly available data collected from a breast phantom (Model 059, CIRS: Tissue Simulation \& Phantom Technology, Norfolk, VA) using an Alpinion E-Cube R12 research US machine (Bothell, WA, USA). The center frequency was 8 MHz and the sampling frequency was 40 MHz. The Young's modulus of the experimental phantom was 20 kPa and contains several inclusions with Young's modulus of higher than 40 kPa. This data is available online at \href{http://code.sonography.ai}{http://code.sonography.ai} in \cite{tehrani2022bi}.

	\textit{In vivo} data was collected at Johns Hopkins hospital from patients with liver cancer during open-surgical RF thermal ablation by a research Antares Siemens system using a VF 10-5 linear array with the sampling frequency of 40 MHz and the center frequency of 6.67 MHz.  The institutional review board approved the study with the consent of the patients. We selected 600 RF frame pairs of this dataset for the training of the networks.

	Two well-known metrics of Contrast to Noise Ratio (CNR) and Strain Ratio (SR) are utilized to evaluate the compared methods. Two Regions of Interest (ROI) are selected to compute these metrics and they can be defined as \cite{ophir1999elastography}:
	\begin{equation}
		\label{Eq:SRCNR}
		CNR = \sqrt{\frac{2(\overline{s}_{b}-\overline{s}_{t})^{2}}{{\sigma _{b}}^{2}+{\sigma _{t}}^{2}}},\quad \quad  SR =\frac{\overline{s}_{t}}{\overline{s}_{b}},
	\end{equation}
	where the subscript $t$ and $b$ denote the target and background ROIs. The SR is only sensitive to the mean ($\overline{s}_{X}$), while CNR depends on both the mean and the standard deviation (${\sigma _{X}}$) of ROIs. For stiff inclusions as the target, higher CNR correlates with better target visibility, and lower SR translates to a higher difference between the target and background strains.

	\subsection{Network architecture and training}
	We employed MPWC-Net++ \cite{tehrani2021mpwc} which has been adapted from PWC-Net-irr \cite{hur2019iterative} for USE. The network architecture with the added known operators is shown in Fig. \ref{fig:net}. The training schedule is similar to \cite{kz2022physically}, known operators are not present in the training and only employed during the test phase. The known operators are added in different pyramid levels. This has the advantage of correcting lateral displacements in different pyramid levels. The known operators are added to the last 3 pyramid levels (there are 5 pyramid levels in this network) since the estimate in the first 2 pyramid levels are not accurate enough and adding the known operators would propagate the error. The hyper-parameters' values of unsupervised training and the known operators are given in Supplementary Materials.

	\vspace{-0.15cm}
	\section{Results and Discussions}
	\subsection{Compared methods}
	kPICTURE is compared to the following methods:
	\begin{itemize}
		
		\item OVERWIND, an optimization-based USE method \cite{mirzaei2019combining}.  
		\item The post-processing method of Guo \textit{et al.} \cite{guo2015pde}, which employs the output of OVERWIND as the initial displacement (OVERWIND+ Guo \textit{et al.}).
		\item PICTURE, which penalize EPR values outside of feasible range \cite{kz2022physically}.    
	\end{itemize}
	We decided to compare with PICTURE instead of sPICTURE \cite{tehrani2022lateral} (PICTURE with self-supervision)
	since self-supervision is not related to the physics of motion. To focus on the effectiveness of
	the known operators, we, therefore, provide a comparison to its corresponding method
	PICTURE. The proposed known operators can be applied to the network trained with sPICTURE method as well. We made the network's weight trained using both PICTURE and sPICTURE methods publicly available online  at \href{http://code.sonography.ai}{http://code.sonography.ai}.  We also employed a similar hyper-parameters and training schedule for experimental phantom and \textit{in vivo} data.

	\subsection{Results and Discussions}
\begin{figure}[t]	
		
		\centering
		\includegraphics[width=0.90\textwidth]{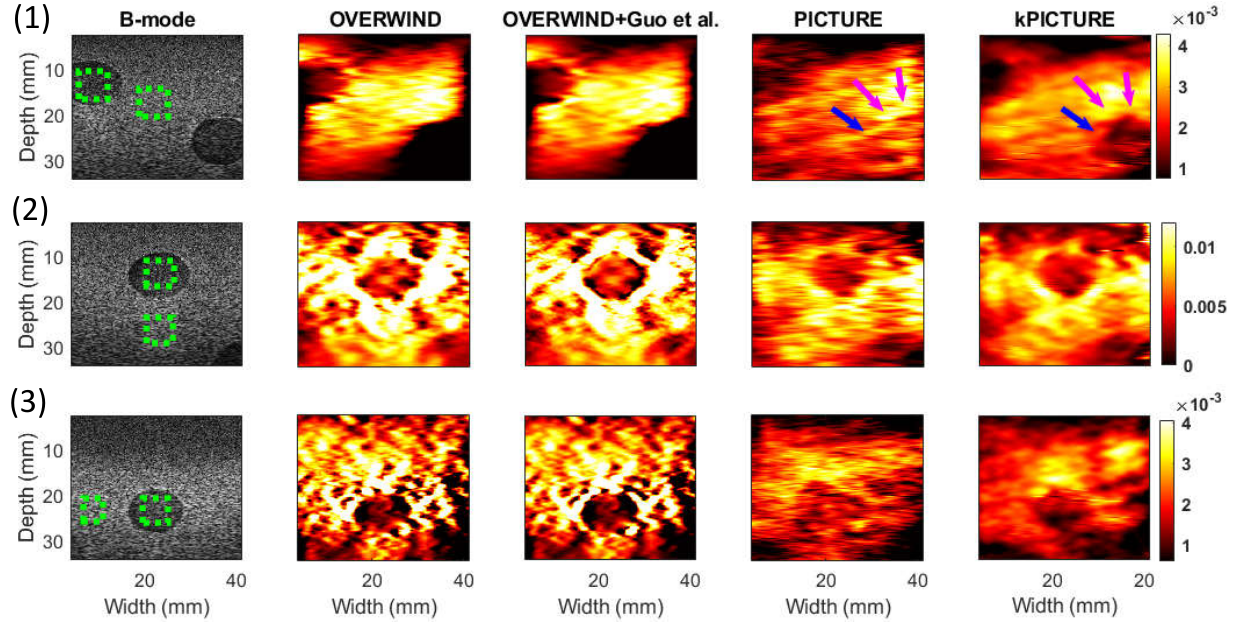}
		\caption{ Lateral strains in the experimental phantom obtained by different methods. The target and background windows for calculation of CNR and SR are marked in the B-mode images. The inclusion on the bottom of sample (1) is highlighted in PICTURE and kPICTURE strain images by purple and blue arrows. The samples 1, 2, and 3 are taken from different locations of the tissue-mimicking breast phantom. Axial strains are available in Supplementary Materials.}
		\label{fig:phantom}
	\end{figure}

	The lateral strains of ultrasound RF data collected from three different locations of the tissue-mimicking breast phantom are depicted in Fig. \ref{fig:phantom}, and the quantitative results are given in Table \ref{tab:phantom}. Visual inspection of Fig. \ref{fig:phantom} denotes that the method proposed by Gou \textit{et al.} \cite{guo2015pde} improves the displacement obtained by OVERWIND. For example, the inclusion borders in sample 2 are much more clearly visible. The strain images obtained by kPICTURE have a much higher quality than those of PICTURE. Furthermore, kPICTURE has the highest quality strain images among the compared methods. For example, the inclusion on the bottom in sample 1 (highlighted by the arrows) is clearly visible in kPICTURE, a substantial improvement over all other methods that do not even show the inclusion.

	\begin{table}[t]
		\caption{Quantitative results of lateral strains for experimental phantoms. Mean and standard deviation ($\pm$) of CNR (higher is better) and SR (lower is better) of lateral strains are reported. The pair marked by $\dagger$ is not statistically significant (\textit{p}-value$>$0.05, using Friedman test). The differences between all other
			numbers are statistically significant (\textit{p}-value$<$0.05). }
		\label{tab:phantom}
		\resizebox{0.999\textwidth}{!}{
			\begin{tabular}{@{}
					>{\columncolor[HTML]{FFFFFF}}c 
					>{\columncolor[HTML]{FFFFFF}}c 
					>{\columncolor[HTML]{EFEFEF}}c 
					>{\columncolor[HTML]{FFFFFF}}c 
					>{\columncolor[HTML]{EFEFEF}}c 
					>{\columncolor[HTML]{FFFFFF}}c
					>{\columncolor[HTML]{EFEFEF}}c 
					>{\columncolor[HTML]{FFFFFF}}c 
					>{\columncolor[HTML]{EFEFEF}}c @{}}
				
				\toprule
				& \multicolumn{2}{c}{\cellcolor[HTML]{FFFFFF}sample (1)} & \multicolumn{2}{c}{\cellcolor[HTML]{FFFFFF}sample (2)} & \multicolumn{2}{c}{\cellcolor[HTML]{FFFFFF}sample (3)} & \multicolumn{2}{c}{\cellcolor[HTML]{FFFFFF}\textit{in vivo data}}\\ 
				& CNR                       & SR                         & CNR                       & SR   & CNR                       & SR & CNR                       & SR                      \\ \midrule
				OVERWIND              & 11.34$\pm$ 1.32    & 0.318$\pm$0.030            & 3.71$\pm$1.07             & 0.505$\pm$0.089     & 3.61 $\pm$0.58            & 0.415$\pm$0.050  &2.07$\pm$0.94 & 0.196$\pm$0.255          \\
				OVERWIND + Gou \textit{et al.} & 13.26 $\pm$ 1.89   & 0.313 $\pm$ 0.029            & 4.28$\pm$1.31     & 0.503$\pm$0.083     & 4.08$\pm$0.62             & \textbf{0.411}$\pm$0.049 & 2.39$\pm$0.89& \textbf{0.170}$\pm$0.233           \\ \midrule
				PICTURE               & 9.037$\pm$0.88     & 0.407$\pm$0.022            & 5.37$\pm$1.33             & 0.449$\pm$0.060$^\dagger$      & 1.63$\pm$0.95             & 0.840$\pm$0.077 &4.36$\pm$1.81 & 0.334$\pm$0.149           \\ 
				kPICTURE              & \textbf{24.40}$\pm$7.02     & \textbf{0.290}$\pm$0.038            & \textbf{7.81}$\pm$1.68       & \textbf{0.446}$\pm$0.056$^\dagger$      & \textbf{5.49}$\pm$2.20    & 0.598$\pm$0.123   & \textbf{5.54}$\pm$2.54&0.504$\pm$0.141        \\ \bottomrule
		\end{tabular}}
	\end{table}

	\begin{figure}	
		
		\centering
		\begin{subfigure}{0.34\textwidth}
			\centering
			\includegraphics[width=1\textwidth]{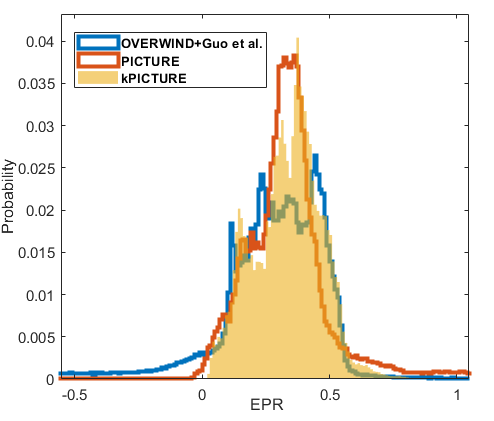}	
			\caption{}	
		\end{subfigure}
		\begin{subfigure}{0.50\textwidth}
			\centering
			\includegraphics[width=1\textwidth]{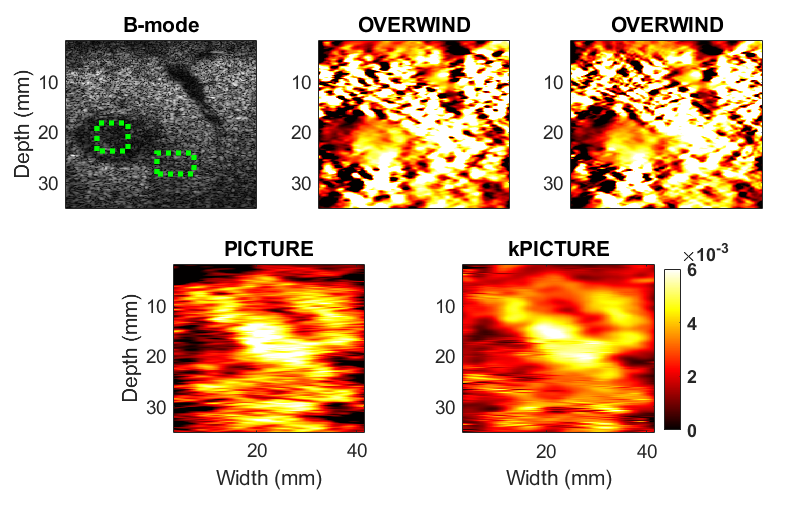}
			\caption{}
		\end{subfigure}
		\caption{The histogram of EPR values for experimental phantom sample 1 (a). The \textit{in vivo} results of the compared methods (b).}
		\label{fig:histogram}
	\end{figure}
	The histograms of EPR values of OVERWIND+Gou \textit{et al.}, PICTURE and kPICTURE are illustrated for the experimental phantom sample (1). To improve visualization, OVERWIND results are not included because the histogram was similar to that of OVERWIND+Gou \textit{et al.}.  Although PICTURE limits the range of EPR using a regularization (Eq \ref{eq:picture}), some EPR values are outside the feasible range. kPICTURE further limits the EPR values; only a small number of samples are outside of the physically plausible range. 
	
	The lateral strain results of \textit{in vivo} data are depicted in Fig. \ref{fig:histogram} (b), and axial strains are given in the Supplementary Materials (the quality of axial strains is high in all methods). While PICTURE may produce an adequate strain image, it still contains noisy regions. On the other hand, kPICTURE delivers exceptionally refined strain images and surpasses the other compared methods. The quantitative results given in table \ref{tab:phantom} also confirm the visual inspection. 
	
	The applied known operators and PICTURE assume that the material is isotropic. Their performance on anisotropic materials can be investigated by experiments on anisotropic tissues such as muscles. Furthermore, 3D imaging data can be collected from 2D arrays to have information in out-of-plane direction to be able to formulate known operators and PICTURE loss for anisotropic tissues.
	
	It should be noted that after incorporating the known operators, the inference time of the network increased from an average of 195 ms to 240 ms (having 10 iterations for algorithm 1 and 100 iterations for algorithm 2).

	\vspace{-0.25cm}
	
	\section{Conclusions}
	In this paper, we proposed to incorporate two known operators inside a USE network. The network is trained by physically inspired constraints specifically designed to tackle the long-standing illusive problem of lateral strain imaging. The proposed operators provide a refinement in each pyramid level of the architecture and substantially improve the lateral strain image quality. Tissue mimicking phantom and \textit{in vivo} results show that the method substantially outperforms previous displacement estimation method in the lateral direction.
	
	\vspace{-0.1cm}
	
	\section{Acknowledgments}
	This research was funded by Natural Sciences and Engineering Research
	Council of Canada (NSERC) Discovery Grant. The Alpinion ultrasound
	machine was purchased using funds from the Dr. Louis G. Johnson
	Foundation.
	
	%
	%
	%
	\bibliographystyle{splncs04}
	\bibliography{IEEEfull}
	%
	
\end{document}